\documentclass{PoS}

\title{Equation of state at finite density in two-flavor QCD with improved Wilson quarks}

\ShortTitle{EOS at finite density in two-flavor QCD}

\author{S. Aoki,$^{a,b}$ S. Ejiri,$^c$ T. Hatsuda,$^d$ N. Ishii,$^e$ \speaker{K. Kanaya},$^a$ Y. Maezawa,$^f$ N. Ukita,$^e$ and T. Umeda$^a$ (WHOT-QCD Collaboration)\\
        \llap{$^a$}Graduate School of Pure and Applied Sciences, Univ. of Tsukuba, Tsukuba 305-8571, Japan\\
        \llap{$^b$}RIKEN BNL Research Center, Brookhaven National Laboratory, Upton, New York 11973, USA\\
        \llap{$^c$}Physics Department, Brookhaven National Laboratory, Upton, New York 11973, USA\\
        \llap{$^d$}Department of Physics, Univ. of Tokyo, Tokyo 113-0033, Japan\\
        \llap{$^e$}Center for Computational Sciences, Univ. of Tsukuba, Tsukuba 305-8577, Japan\\
        \llap{$^f$}En'yo Laboratory, Nishina Accelerator Research Center, RIKEN, Wako 351-0198, Japan\\
        E-mail: \email{kanaya@ccs.tsukuba.ac.jp}}

\abstract{We study the equation of state in two-flavor QCD at finite temperature and density.  Simulations are made with the RG-improved gluon action and the clover-improved Wilson quark action.  Along the lines of constant physics for  $m_{\rm PS}/m_{\rm V} = 0.65$ and 0.80, we compute the derivatives of the quark determinant with respect to the quark chemical potential $\mu_q$ up to the fourth order at $\mu_q=0$.  We adopt several improvement techniques in the evaluation.  We study thermodynamic quantities and quark number susceptibilities at finite $\mu_q$ using these derivatives. We find enhancement of the quark number susceptibility at finite $\mu_q$, in accordance with previous observations using staggered-type quarks.  This suggests the existence of a nearby critical point.}

\FullConference{The XXVI International Symposium on Lattice Field Theory\\
		 July 14-19 2008\\
		 Williamsburg, Virginia, USA}

\begin{document}

\section{Introduction}

Finite density QCD has been studied on the lattice mainly using staggered-type quarks \cite{EjiriLat08}. 
However, because the expected O(4) universality of the deconfining transition in two-flavor QCD has not been confirmed with staggered-type quarks, the results may contain sizable lattice artifacts. 
Therefore, we need to crosscheck the results with different lattice actions. 
We study it with Wilson-type quarks, with which the O(4) scaling has been confirmed \cite{O4,CP-PACS00}.

Because Wilson-type quarks are numerically more intensive, we have to adopt/develop several improvement techniques.  
We apply a hybrid method of Taylor expansion and spectral reweighting, and develop a couple of improvement tricks.

\section{Formulation}

We extend the study of two-flavor QCD by the CP-PACS Collaboration at vanishing chemical potential $\mu_q=0$ \cite{CP-PACS00} to finite densities.  
Preliminary reports of this study have been presented at Lattice 2006 and 2007 conferences \cite{prelim}.

\begin{figure}
\centerline{
\includegraphics[width=.33\textwidth]{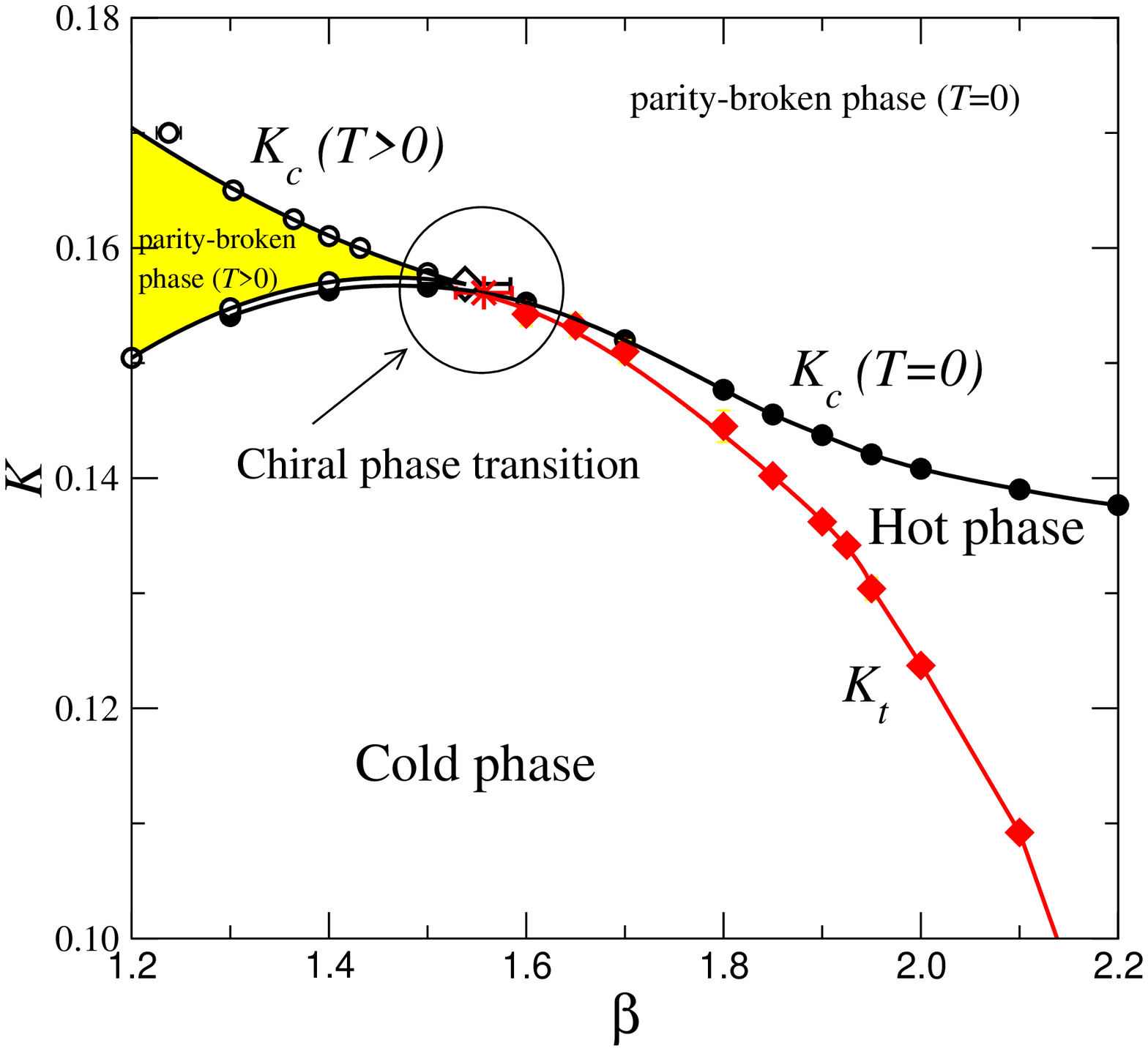}
\hspace{8mm}
\includegraphics[width=.4\textwidth]{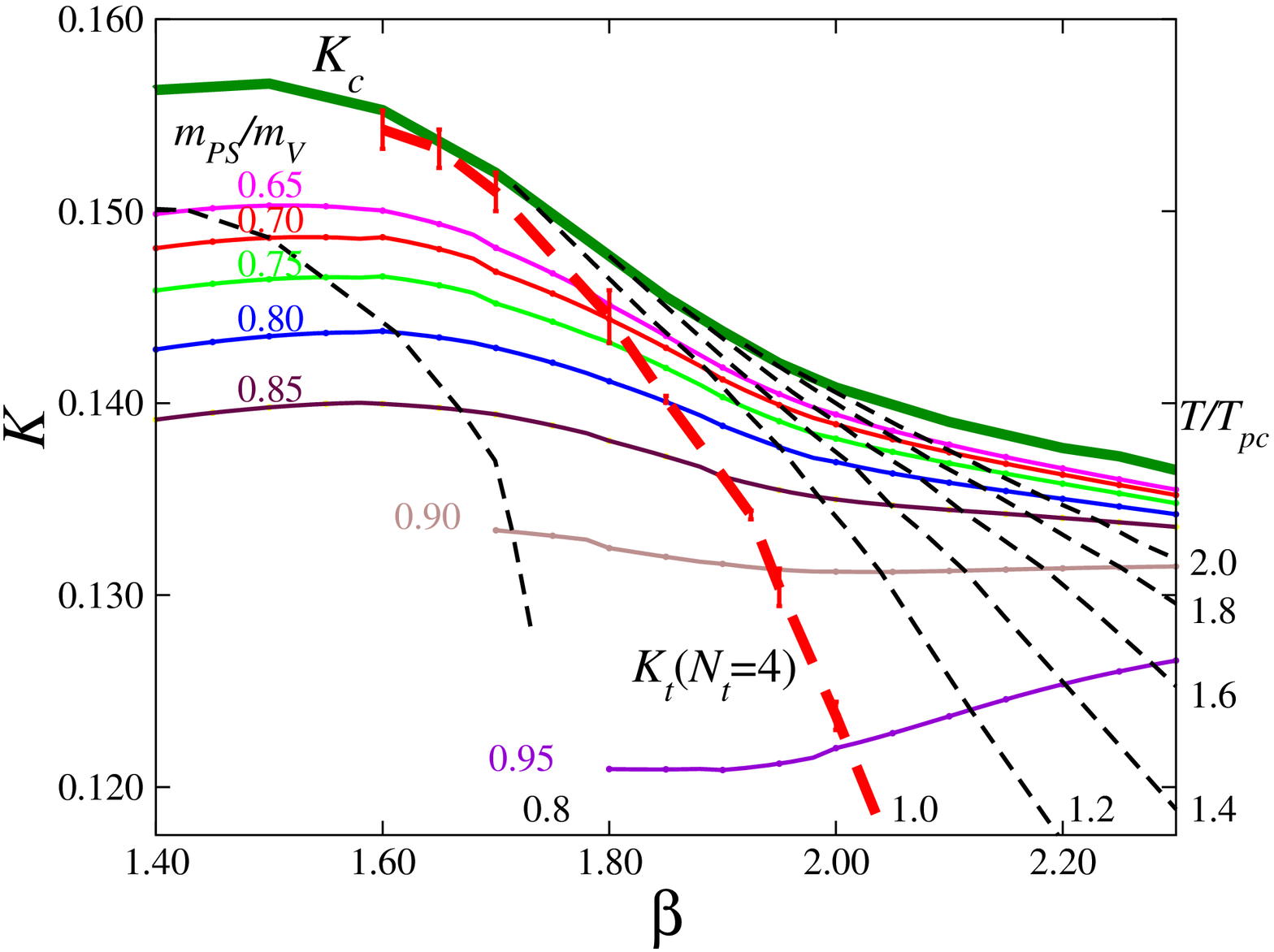}
}
\caption{Phase diagram and lines of constant physics at $\mu=0$ for two-flavor QCD with improved Wilson quarks and Iwasaki glue at $N_t=4$ \cite{CP-PACS00,param}.}
\label{fig1}
\end{figure}

We adopt the clover-improved Wilson quark action and the RG-improved Iwasaki gauge action defined by 
$S = S_g + S_q$ with
\begin{eqnarray}
  S_g &=& 
  -{\beta}\sum_x\left(
   c_0\!\sum_{\mu<\nu;\mu,\nu=1}^{4}\!\!W_{\mu\nu}^{1\times1}(x) 
   +c_1\!\sum_{\mu\ne\nu;\mu,\nu=1}^{4}\!\!W_{\mu\nu}^{1\times2}(x)\right), \;\;\;
  S_q = \sum_{f=1}^2\sum_{x,y}\,\bar{q}_x^f M_{x,y}\,q_y^f,
  \label{eq:action}
\end{eqnarray}
where $\beta=6/g^2$, $c_1=-0.331$, $c_0=1-8c_1$, and
\begin{eqnarray}
 M_{x,y} &=& \delta_{xy}
   -{K}\sum_{i=1}^3 \{(1-\gamma_{i})U_{x,i}\delta_{x+\hat{i},y}
    +(1+\gamma_{i})U_{x,i}^{\dagger}\delta_{x,y+\hat{i}}\}
       \nonumber \\ &&
   -{K} \{e^{\mu}(1-\gamma_{4})U_{x,4}\delta_{x+\hat{4},y}
    +e^{-\mu}(1+\gamma_{4})U_{x,4}^{\dagger}\delta_{x,y+\hat{4}}\}
   -\delta_{xy}{c_{SW}}{K}\sum_{\mu<\nu}\sigma_{\mu\nu}F_{\mu\nu}.
\label{eq:fermact}
\end{eqnarray}
where $\mu \equiv \mu_q a$ and $F_{\mu\nu}$ is the lattice field strength in terms of 
the standard clover-shaped combination of gauge links.
For the clover coefficient $c_{SW}$, we adopt a mean field value using
$W^{1\times 1}$ calculated in the one-loop perturbation theory: 
$ {c_{SW}}=(W^{1\times 1})^{-3/4}=(1-0.8412\beta^{-1})^{-3/4}$ \cite{CP-PACS00}.

The lattice size is $N_s^3\times N_t = N_{\rm site} =16^3\times4$.
We carry out simulations along the lines of constant physics (LOCs) for $m_{PS}/m_V = 0.65$ and 0.80 at $\mu_q = 0$ \cite{param} shown in Fig.\ref{fig1}.
For definiteness, we refer the pseudocritical temperature $T_{pc}$ at $\mu=0$ as $T_0$ in the followings.
Runs are carried out in the range $\beta=1.50$--2.40 at thirteen values of
$T/T_0\sim 0.82$--4.0 for $m_{\rm PS}/m_{\rm V}=0.65$ and twelve values of
$T/T_0\sim 0.76$--3.0 for $m_{\rm PS}/m_{\rm V}=0.80$.
The number of trajectories for each run after thermalization is
5000--6000. We measure physical quantities at every 10 trajectories.

\section{Taylor expansion up to $O(\mu_q^4)$}

We expand the grand canonical potential $\omega = p/T^4$ in terms of $\mu_q/T$ at $\mu_q=0$:
\begin{eqnarray}
\omega = \frac{1}{VT^3}\ln{\cal Z} = \sum_{n=0}^{\infty} c_n(T) \left(\frac{\mu_q}{T}\right)^n,
\;\;\;
c_n(T) = \frac{1}{n!} \frac{N_t^{3}}{N_s^3} \left.
\frac{\partial^n \ln{\cal Z}}{\partial(\mu_q/T)^n} \right|_{\mu_q=0},
\end{eqnarray}
where $V=(N_s a)^3$ is the system volume and ${\cal Z}$ is the partition function.
We expand up to $(\mu_q/T)^4$.
The relevant coefficients are given by
\begin{eqnarray}
&&
c_2 = \frac{N_t}{2N_s^3} {\cal A}_2 ,
\hspace{3mm}
c_4 = \frac{1}{4! N_s^3 N_t} ({\cal A}_4 -3 {\cal A}_2^2) , 
\\
&&
{\cal A}_2 =
\left\langle {\cal D}_2 \right\rangle 
+\left\langle {\cal D}_1^2 \right\rangle, 
\hspace{3mm}
{\cal A}_4 =
\left\langle {\cal D}_4 \right\rangle 
+4\left\langle {\cal D}_3 {\cal D}_1 \right\rangle 
+3\left\langle {\cal D}_2^2 \right\rangle 
+6\left\langle {\cal D}_2 {\cal D}_1^2 \right\rangle 
+\left\langle {\cal D}_1^4 \right\rangle , 
\nonumber
\end{eqnarray}
where
$
{\cal D}_n = N_f \left(\frac{\partial}{\partial \mu}\right)^n \ln \det M
$
is calculated as 
\[
{\cal D}_1
= N_f {\rm tr} \left( M^{-1} \frac{\partial M}{\partial \mu} \right),
\hspace{3mm}
{\cal D}_2
= N_f \left[ {\rm tr} \left( M^{-1} \frac{\partial^2 M}{\partial \mu^2} \right)
 - {\rm tr} \left( M^{-1} \frac{\partial M}{\partial \mu}
                   M^{-1} \frac{\partial M}{\partial \mu} \right) \right] , 
\hspace{3mm}{\rm etc.}
\]
We calculate similar coefficients for the expansion in terms of the isospin chemical potential $\mu_I$ too.

\begin{figure}
\centerline{
\includegraphics[width=.32\textwidth]{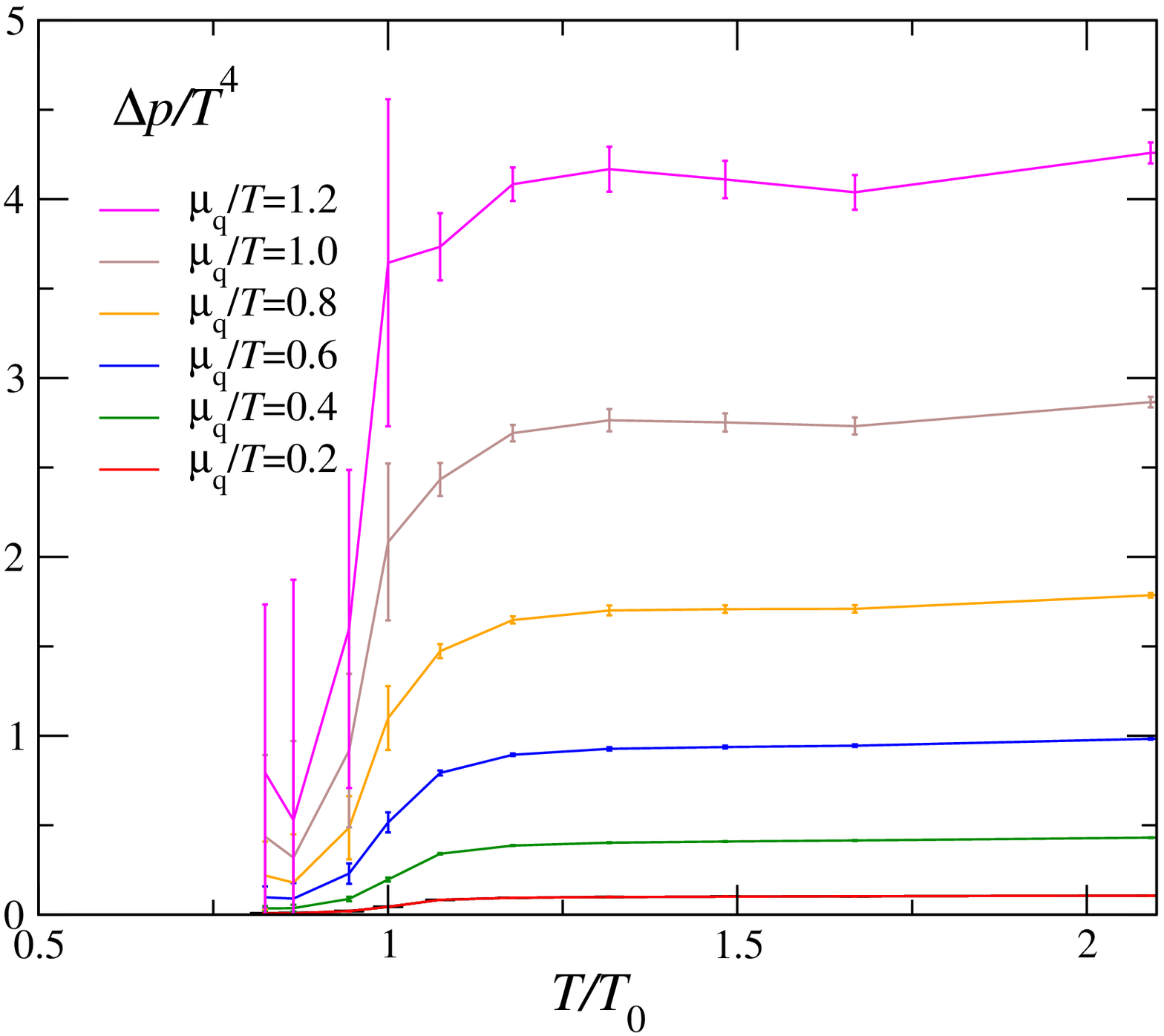}
\hspace{10mm}
\includegraphics[width=.32\textwidth]{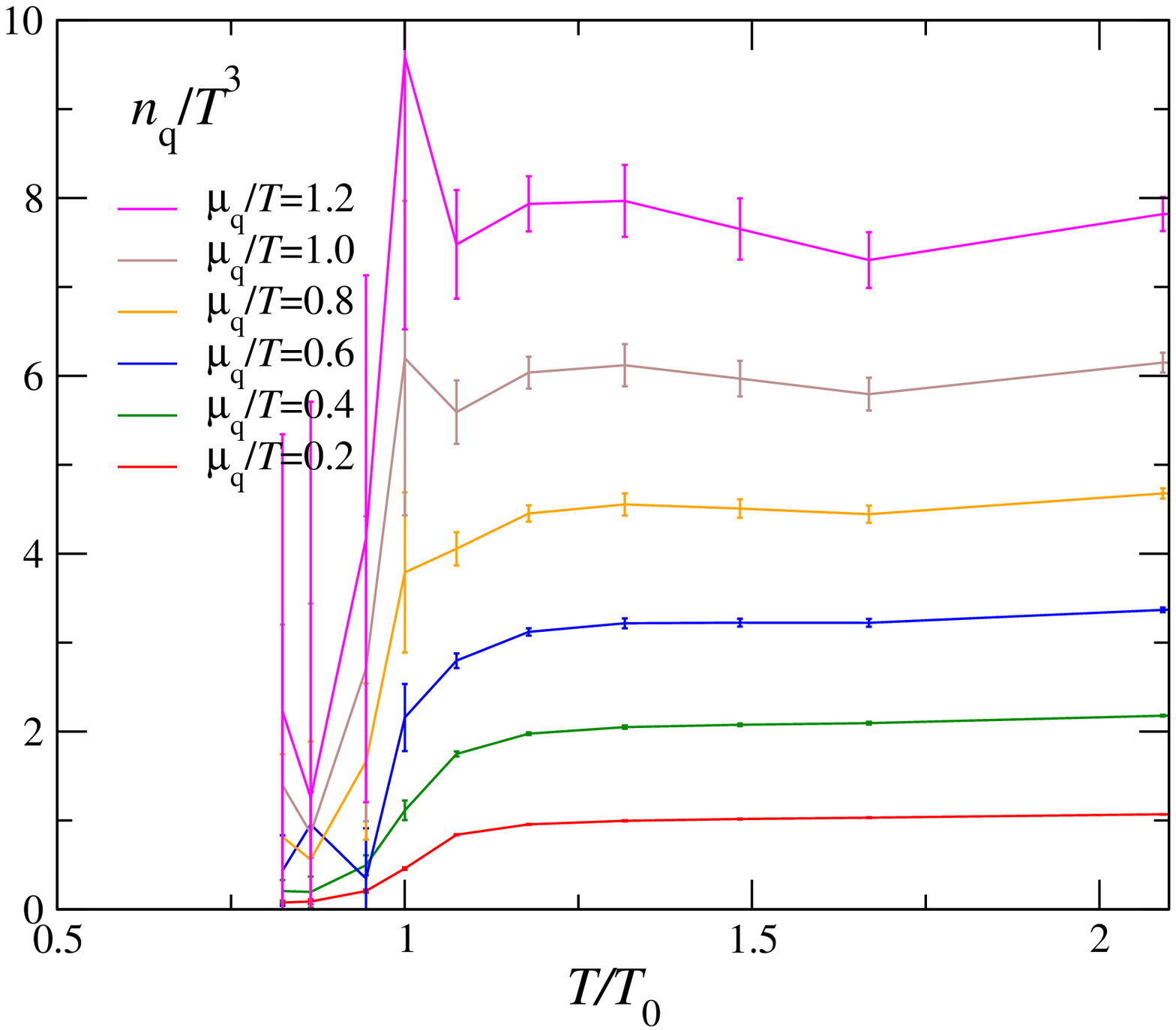}
}
\centerline{
\includegraphics[width=.32\textwidth]{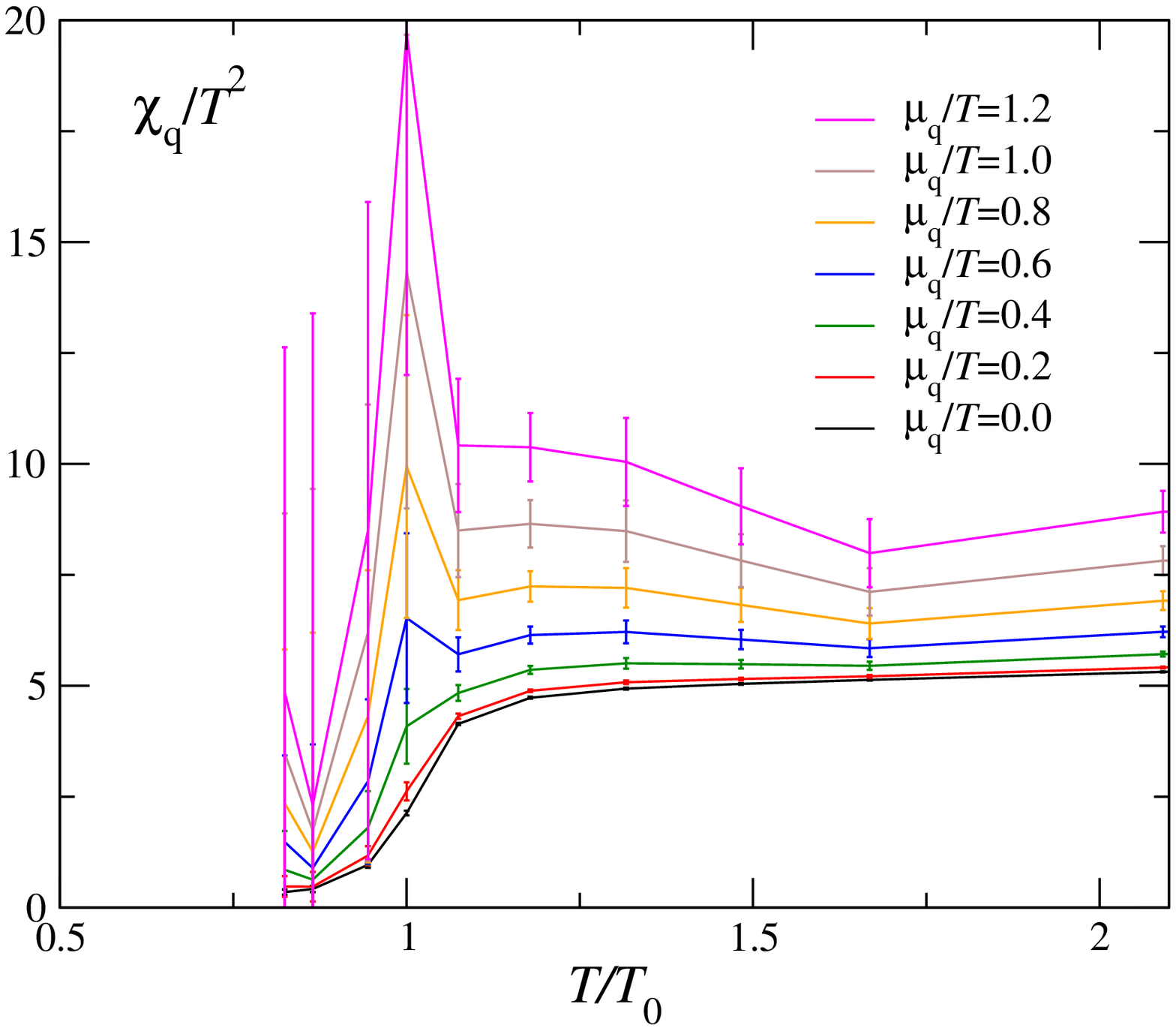}
\hspace{10mm}
\includegraphics[width=.32\textwidth]{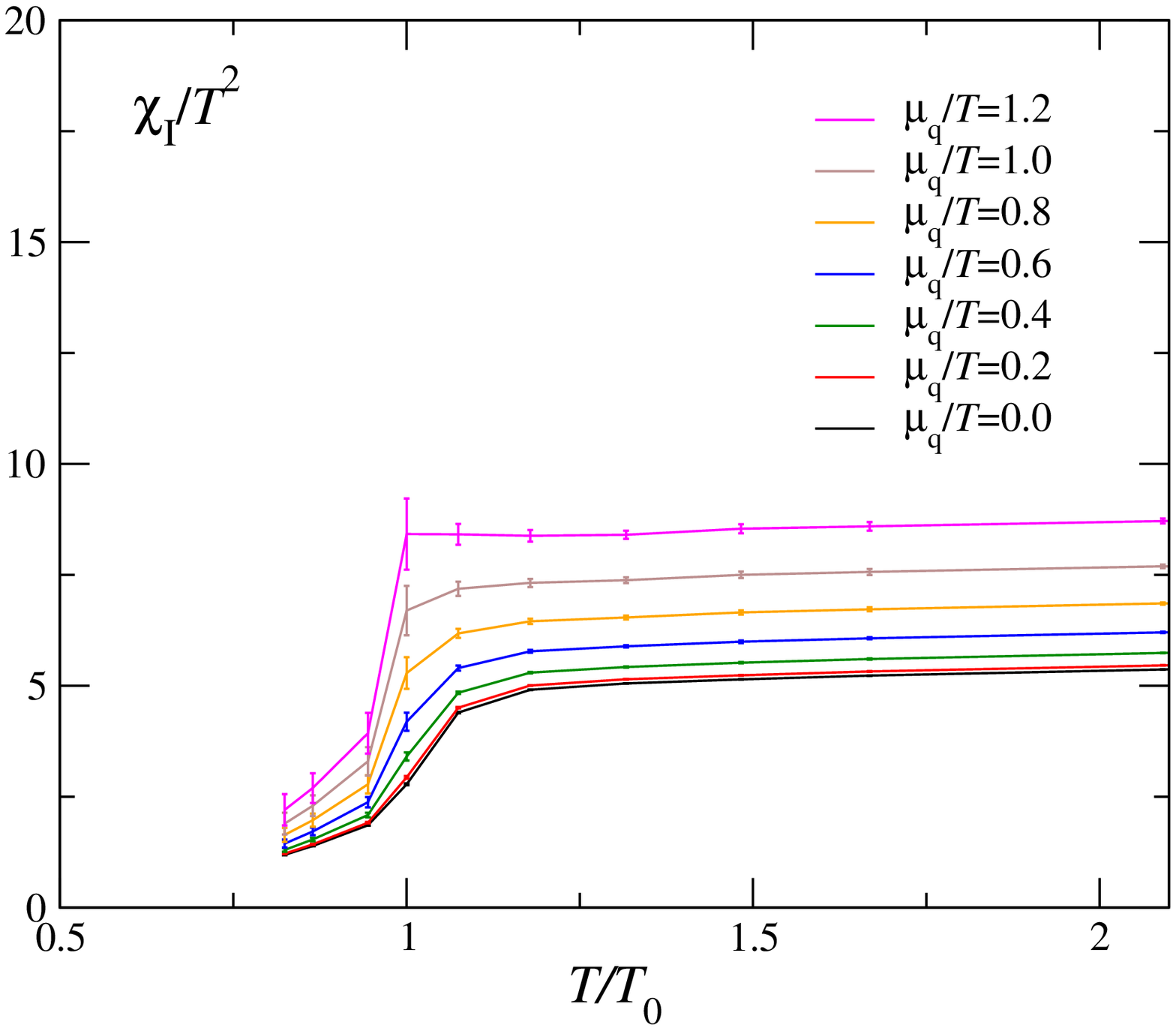}
}
\caption{Results of the Taylor expansion method up to the fourth order of chemical potentials for $m_{\rm PS}/m_{\rm V} = 0.65$. 
$T_0$ is the pseudocritical temperature at $\mu=0$ for the same LOC.}
\label{figTaylor}
\end{figure}

We evaluate the traces with the random noise method. 
In the evaluation, we apply the following two improvements: 
(i) Because the elements off-diagonal
in color and spin indices are not suppressed by $|x-y|$ with Wilson-type quarks,
the number of the same-magnitude off-diagonal elements in the quark matrix is 11
times larger than the diagonal one. This is different from the case of staggered-type
quarks, in which off-diagonal elements in spin indices are slightly suppressed by the
spatial offset. Because a large number of noises is required to pick up a signal from
data with $S/N = 1/11$, we decide not to apply the noise method for color and spin
indices and generate noise vectors only for spatial indices, i.e. we repeat the inversion
of $M$ for each color and spin indices. 
(ii) We find that the dominant errors are from terms containing ${\cal D}_1$.
Therefore, we adopt 10-40 times more noise vectors for ${\cal D}_1$. while we generate only 10 noise vectors for other traces.

Results for the pressure difference $\Delta p = p(\mu_q) - p(0)$, quark number density ($n_q$), 
and quark number as well as isospin susceptibilities ($\chi_q$ and $\chi_I$, respectively) are shown in Fig\ref{figTaylor} for $m_{\rm PS}/m_{\rm V} = 0.65$.
We observe larger enhancement in the quark number fluctuations near $T_0$ with increasing $\mu_q$, 
however such an enhancement around $T_0$ is not shown 
in the isospin fluctuations.
Because the statistical errors in $n/T^3$ and $\chi_q/T^2$ are large, 
further studies are needed increasing the statistics for more precise 
arguments with this approach.

\section{Improvement:  A Hybrid Method}

In the previous section, the Taylor expansion is limited up to the fourth order of the chemical potential,
because the evaluation of ${\cal D}_n$ with $n > 4$ is compuationally demanding.
To improve the calculation, we need to estimate $c_n$ at larger $n$.
Here, we note that ${\cal D}_n=0$ at $n > 4$ for the free quark case. 
Therefore, at high temperatures, we may approximate ${\cal D}_n=0$ for $n > 4$ in the evaluation of $c_n$ at $n > 4$. 
This approximation corresponds to a hybrid reweighting method in which the grand canonical potential is approximated by a truncated Taylor expansion
\begin{eqnarray}
\omega(T,\mu) 
& \approx &
\frac{1}{VT^3} \ln {\cal Z}(T,0) 
+ \frac{1}{VT^3} \ln \left\langle \exp \left[ 
\sum_{n=1}^{N_{\rm max}} {\cal D}_n \, \mu^n 
\right] \right\rangle_{(\mu=0)}, 
\label{eq:pranpotap} 
\end{eqnarray}
with $N_{\rm max}=4$.
Here, $\langle \cdots \rangle_{(\mu=0)}$ is the average over configurations at $\mu=0$.
We then have
\begin{equation}
{\cal Z}(T,\mu) \approx {\cal Z}(T,0) \left\langle e^{F(\mu)} 
e^{ i\theta(\mu)} \right\rangle_{(\mu=0)}
\end{equation}
with
\begin{eqnarray}
\theta (\mu) 
&\equiv & N_{\rm f} {\rm Im} [\ln \det M(\mu)]
\approx {\rm Im} {\cal D}_{1} \mu + \frac{1}{3!} {\rm Im} {\cal D}_{3} \mu^{3},
\\
F (\mu) 
&\equiv & N_{\rm f} {\rm Re} \left[ \ln \left( \frac{\det M(\mu)}{\det M(0)} \right) \right]
\approx \frac{1}{2!} {\rm Re} {\cal D}_{2} \mu^{2} + \frac{1}{4!} {\rm Re} {\cal D}_{4} \mu^{4}.
\end{eqnarray}

This kind of hybrid method was first tested in 
\cite{BS05} with two flavors of staggered quarks. 
In that study, it turned out to be difficult to control statistical errors at $\mu_q$ larger than $O(T)$.
This is due to the sign problem at finite $\mu_q$.

\begin{figure}
\centerline{
\includegraphics[width=.32\textwidth]{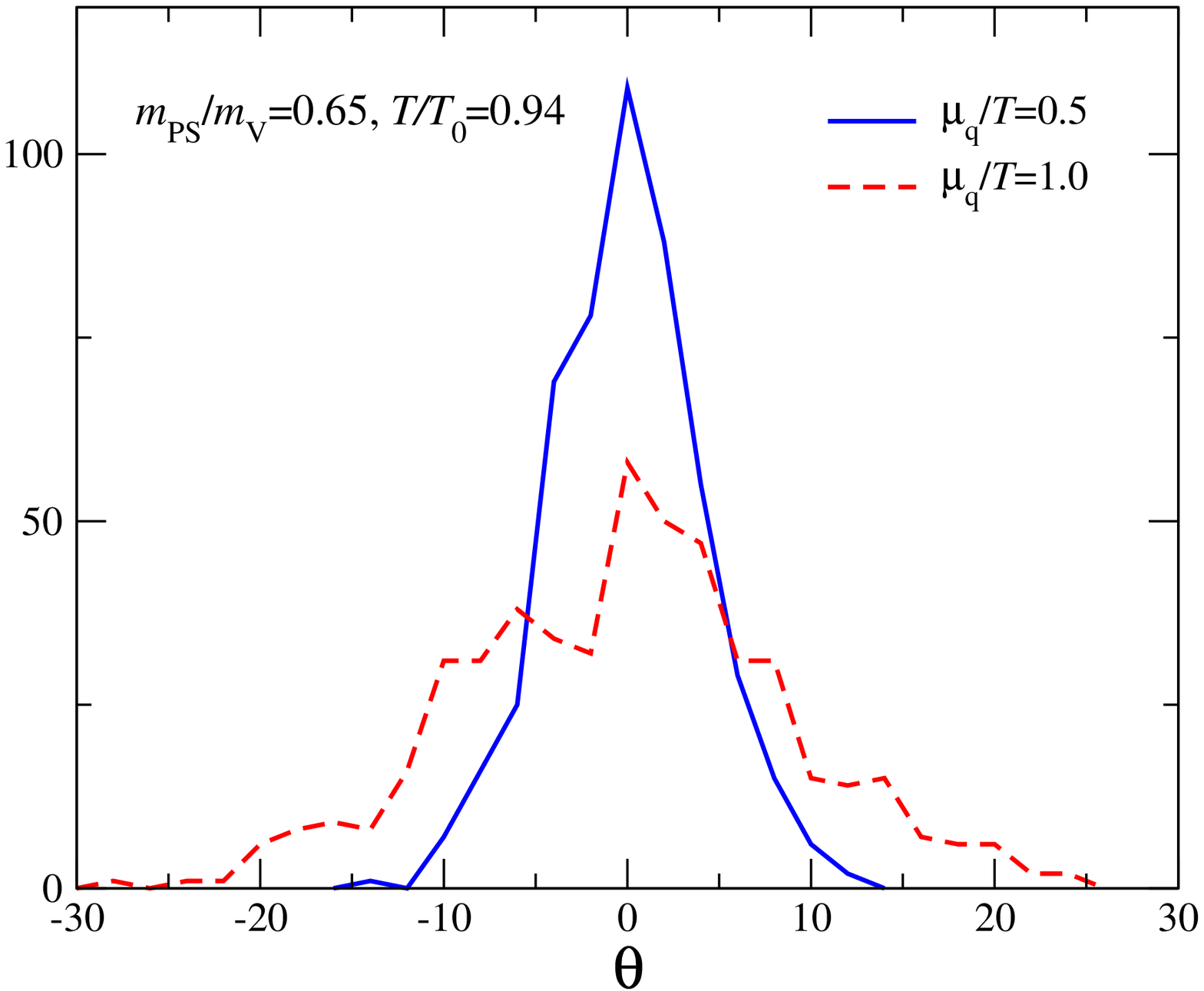}
\hspace{10mm}
\includegraphics[width=.32\textwidth]{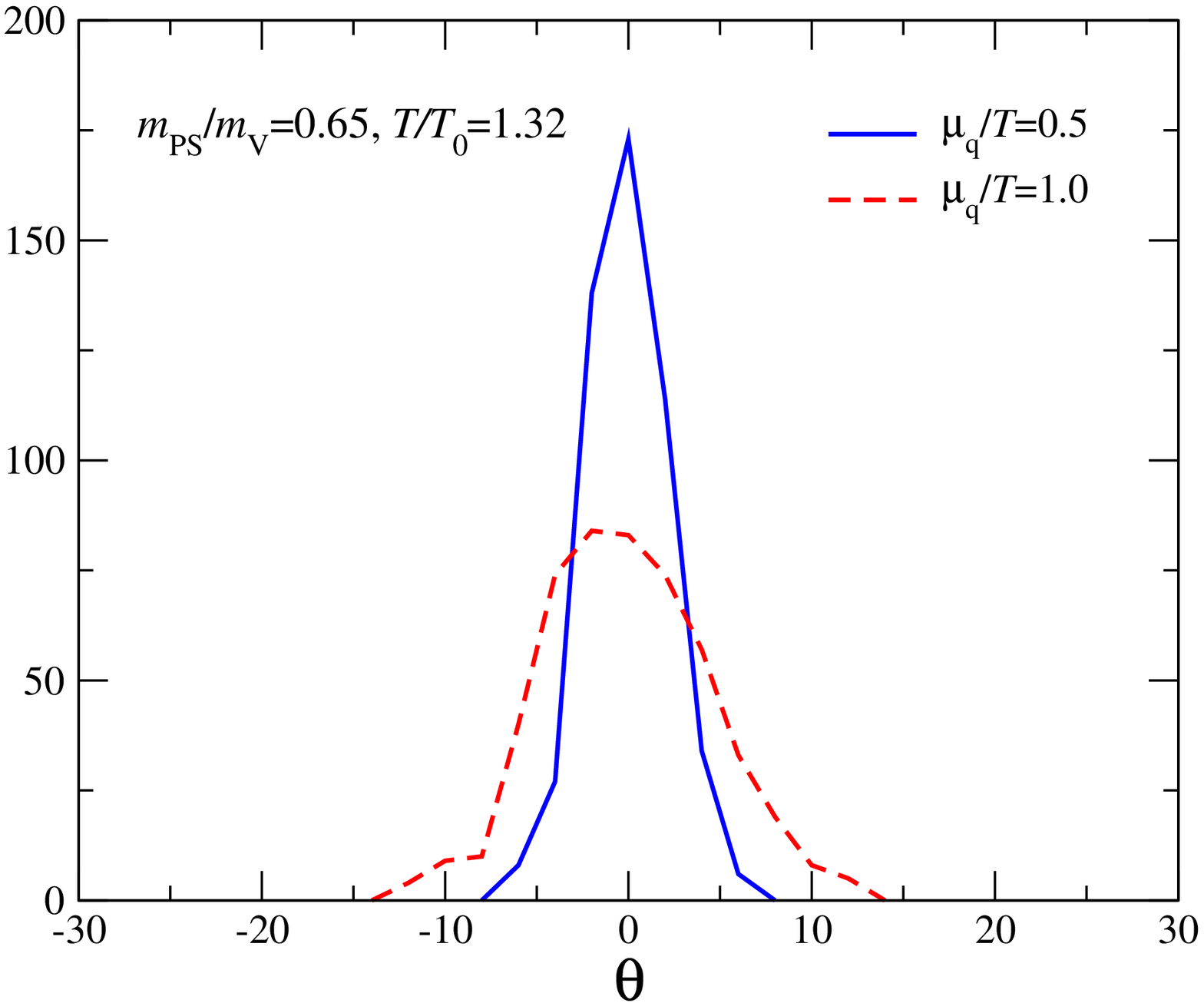}
}
\caption{Typical $\theta$-distribution of two-flavor QCD at $\mu_q=0$ for $m_{\rm PS}/m_{\rm V} = 0.65$. }
\label{figGaussian}
\end{figure}

\begin{figure}
\centerline{
\includegraphics[width=.32\textwidth]{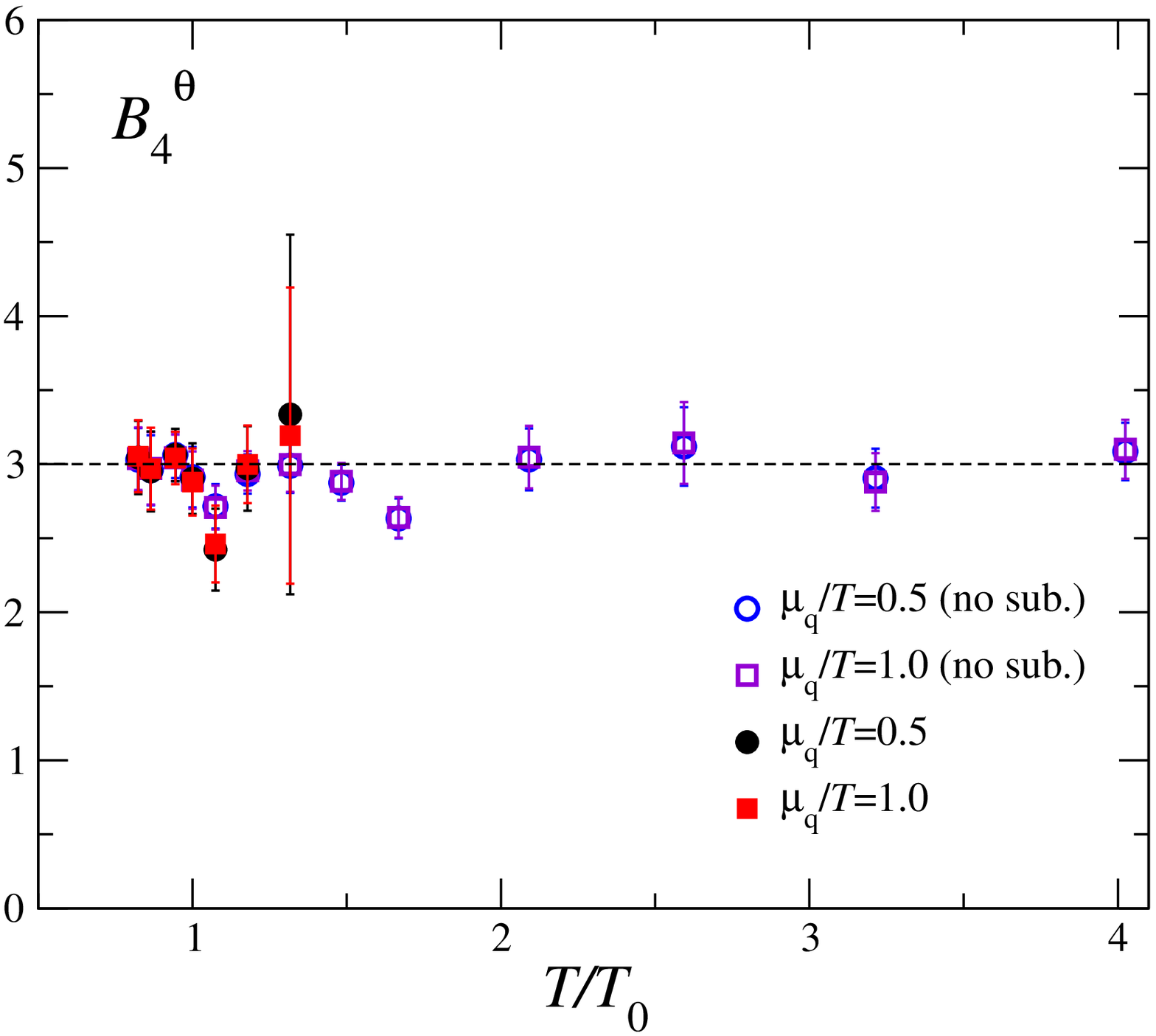}
\hspace{10mm}
\includegraphics[width=.32\textwidth]{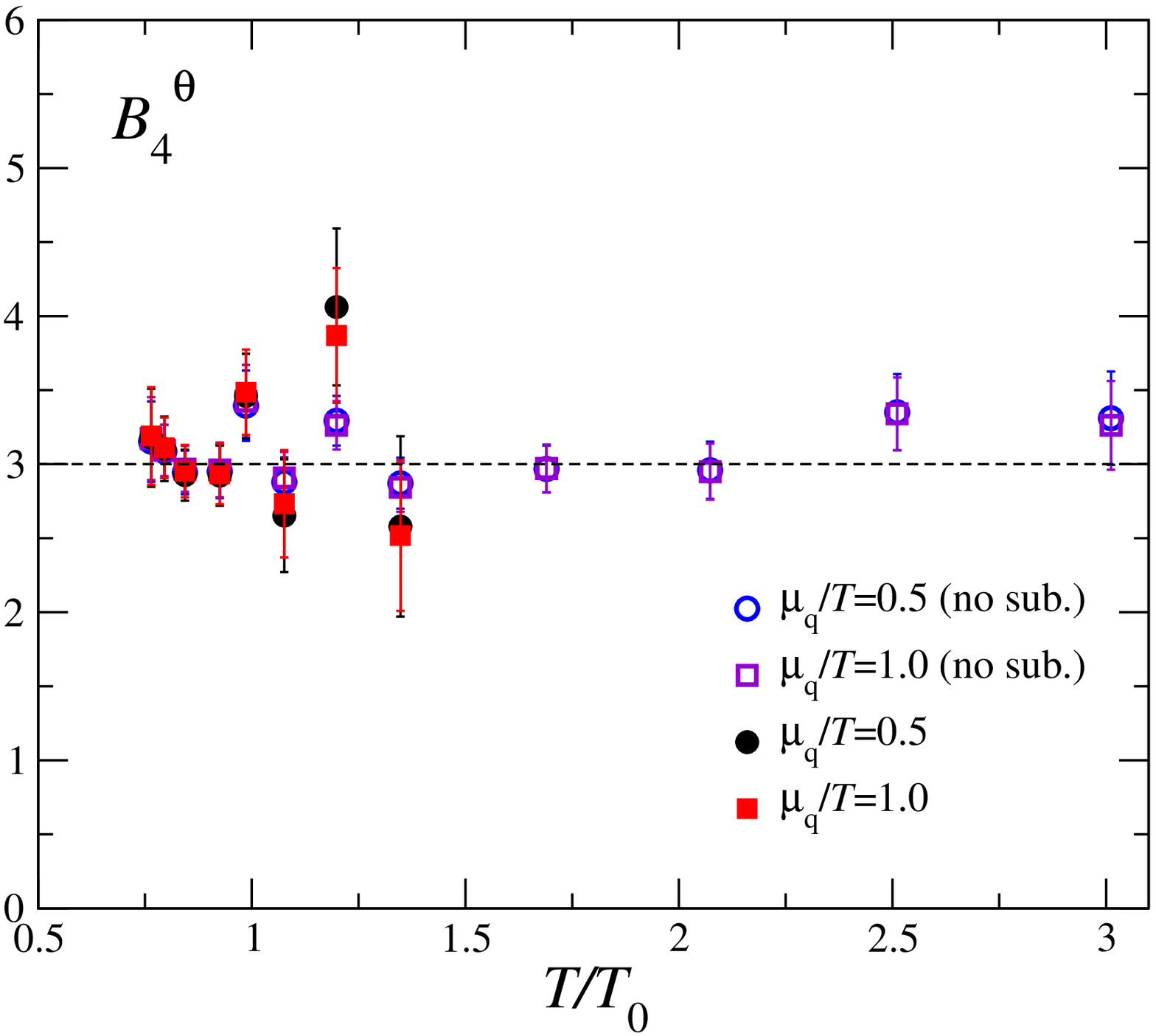}
}
\caption{Binder cumulant test of $\theta$-distribution for $m_{\rm PS}/m_{\rm V} = 0.65$ (left) and 0.80 (right). 
Data labeled as ``no sub.'' are the results of approximate estimation of $\theta^2$ and $\theta^4$ without subtracting the contribution of the same noise vectors. 
Data without  ``no sub.'' are the correct estimates.}
\label{figBinder}
\end{figure}

In a previous study with staggered quarks, SE noted that the $\theta$-distribution at $\mu=0$ is well described by a Gaussian form, and showed that this fact can be used to carry out the $\theta$-averaging with small errors \cite{ejiri08}.
Furthermore, we can argue that, because $\theta = \cal{O}(\mu)$ at small $\mu$, deviations from the Gaussian distribution do not affect the terms up to $\mu^4$ \cite{ejiri08-2,paperWHOT}. 
We find that our data are also well Gaussian (see Fig.\ref{figGaussian} for typical examples).
A convenient test of Gaussian distributions is provided by the Binder cumulant
$
B_4^{\theta} = \left\langle \theta^4 \right\rangle / \left\langle \theta^2 \right\rangle^2
$.
If the distribution is Gaussian, we expect $B_4^{\theta} = 3$.
In Fig.\ref{figBinder}, we show that all of our distributions are approximately consistent with $B_4^{\theta} = 3$.

With Gaussian $\theta$-distribution, we can carry out the $\theta$-averaging, and the task is reduced to evaluate 
\begin{equation}
{\cal Z}(T,\mu) \approx {\cal Z}(T,0) \left\langle e^F 
e^{ -\frac{1}{2} \left\langle \theta^2 \right\rangle_{F}} \right\rangle_{(\mu=0)},
\end{equation}
with $\left\langle \theta^2 \right\rangle_{F}$ the mean $\theta^2$ for given fixed $F$.

The remaining $F$-averaging is again challenging because the factor $e^{F(\mu)}$ can easily shift the central contribution for the average to a statistically poor region of $F$.
At small $\mu$, this problem can be largely resolved by shifting $\beta$ with $\mu$ such that 
the fluctuation in $e^{F(\mu)}$ is compensated by that in the gauge action, 
because $F$ is sensitively correlated with the gauge action $S_g$. 
More concretely, we adopt the reweighting method to shift $\beta$ and calculate an optimal $\beta$ by minimizing the fluctuation 
\[
\left\langle \left(e^F e^{ -\frac{1}{2} \left\langle \theta^2 \right\rangle_{F}}
e^{6 N_{\rm site} (\beta -\beta_0) P} - 
\left\langle e^F e^{ -\frac{1}{2} \left\langle \theta^2 \right\rangle_{F}} 
e^{6 N_{\rm site} (\beta -\beta_0) P} \right\rangle \right)^2 
\right\rangle 
\]
where $P = S_g/(6 N_{\rm site} \beta)$ is the generalized plaquette. 
The shifts in $\beta$ turn out to be less than about 0.03 in our study.
Since these are negligible in Fig.\ref{fig1}, we disregard the resulting small deviation from the line of constant physics, and simply translate the shifts in $\beta$ to shifts in $T$ for the final plots.

\begin{figure}
\centerline{
\includegraphics[width=.32\textwidth]{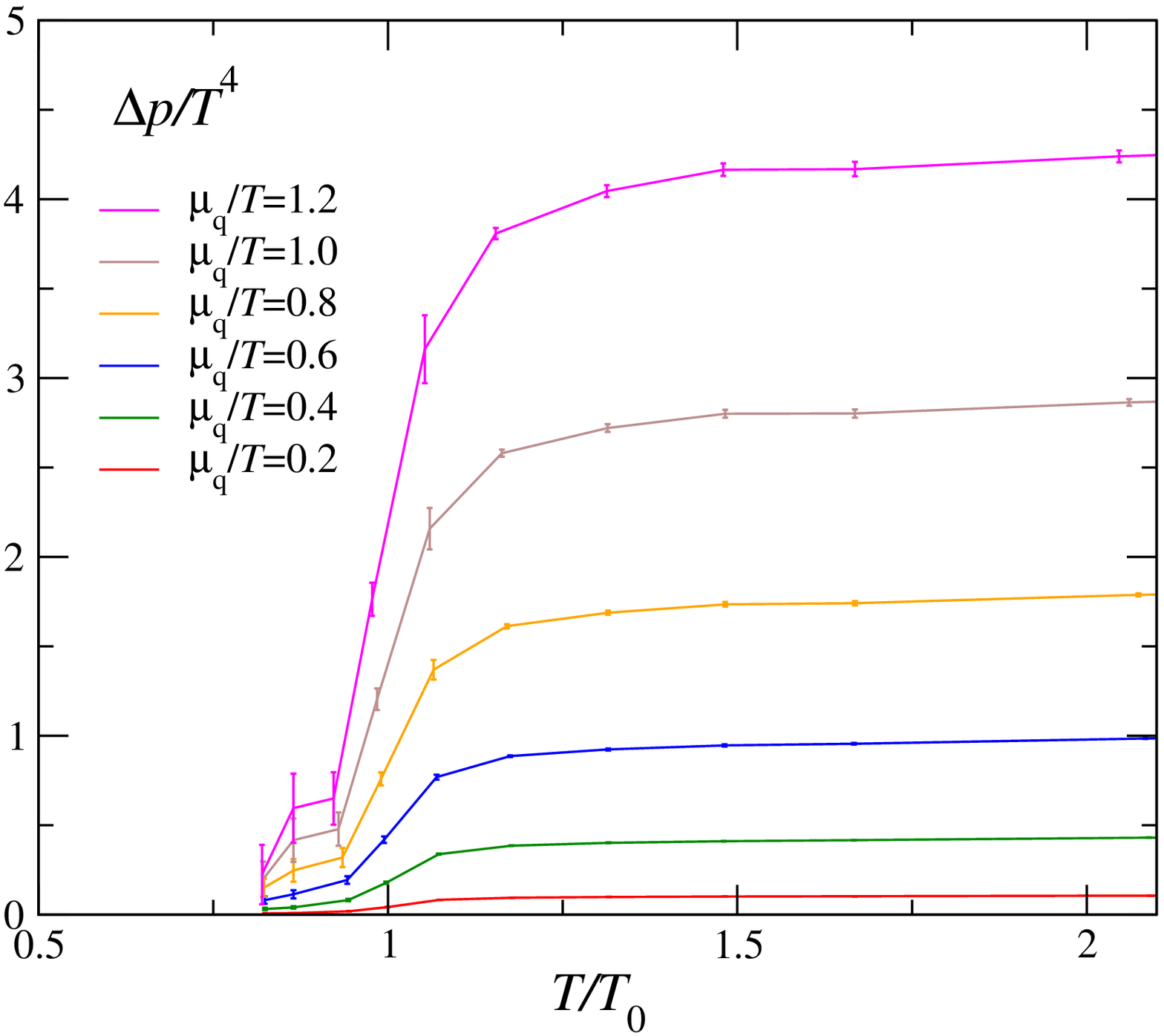}
\hspace{1mm}
\includegraphics[width=.32\textwidth]{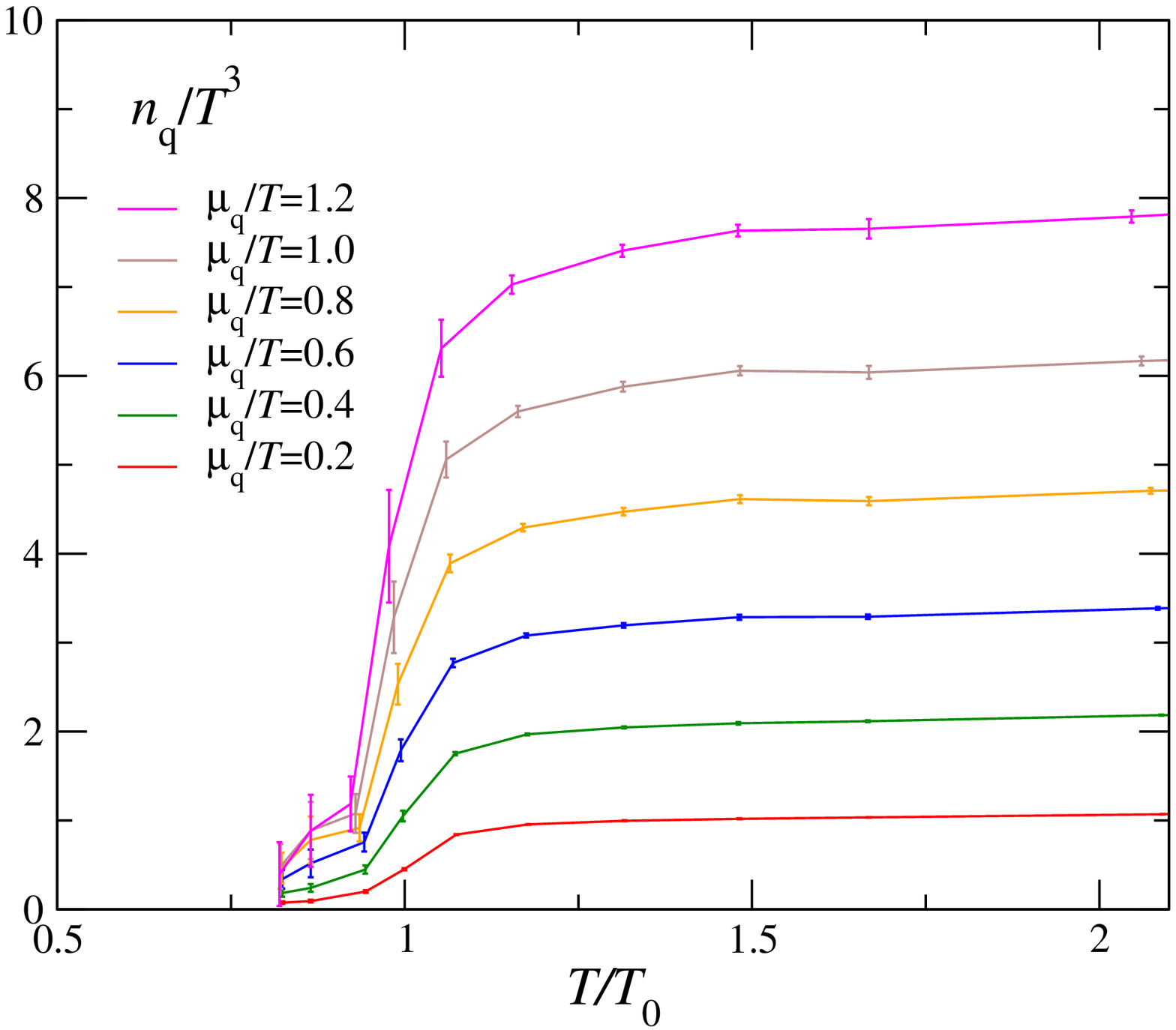}
\hspace{1mm}
\includegraphics[width=.32\textwidth]{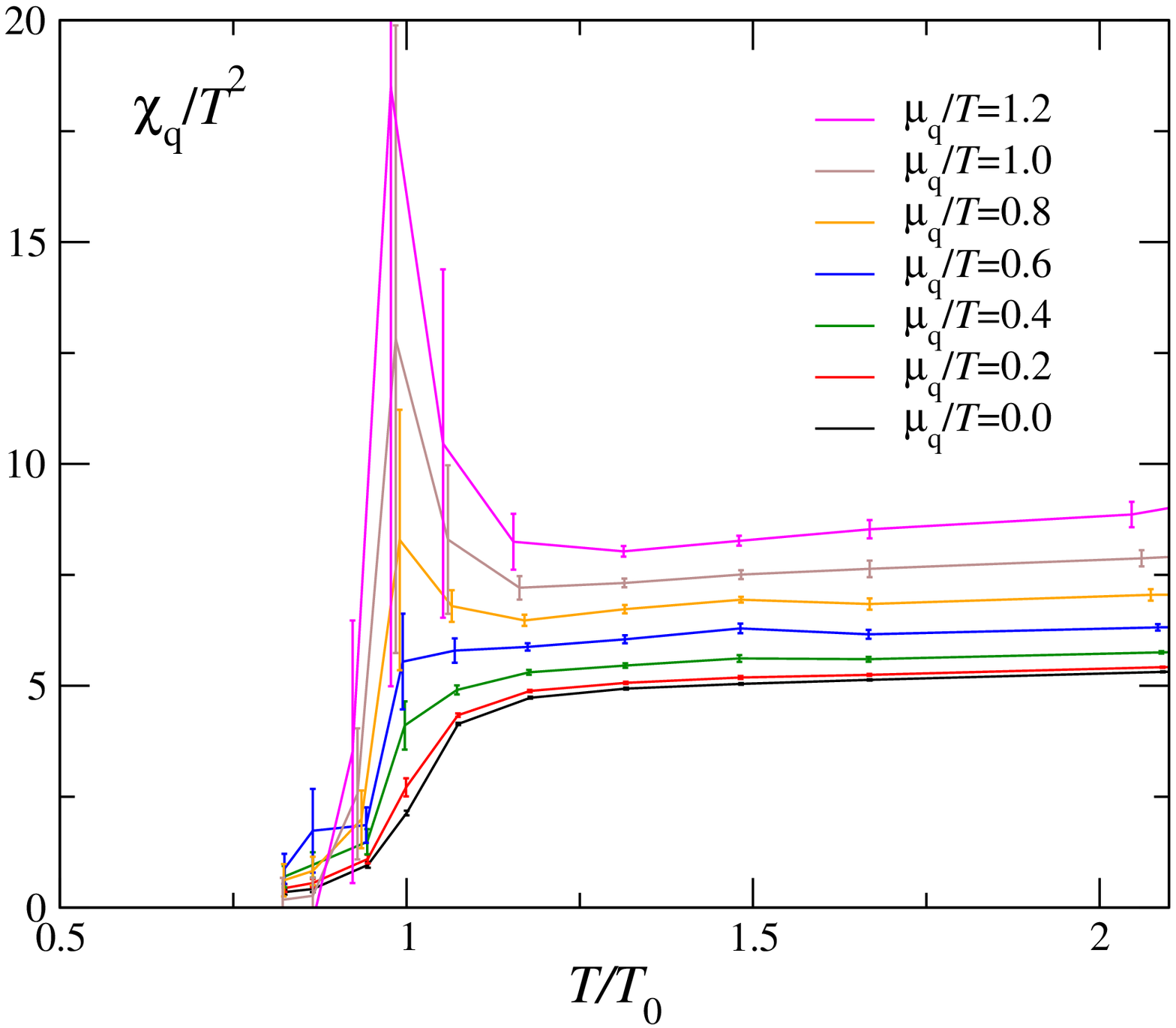}
}
\caption{Results of EOS using the hybrid method for $m_{\rm PS}/m_{\rm V} = 0.65$.}
\label{figHybrid65}
\end{figure}

\begin{figure}
\centerline{
\includegraphics[width=.32\textwidth]{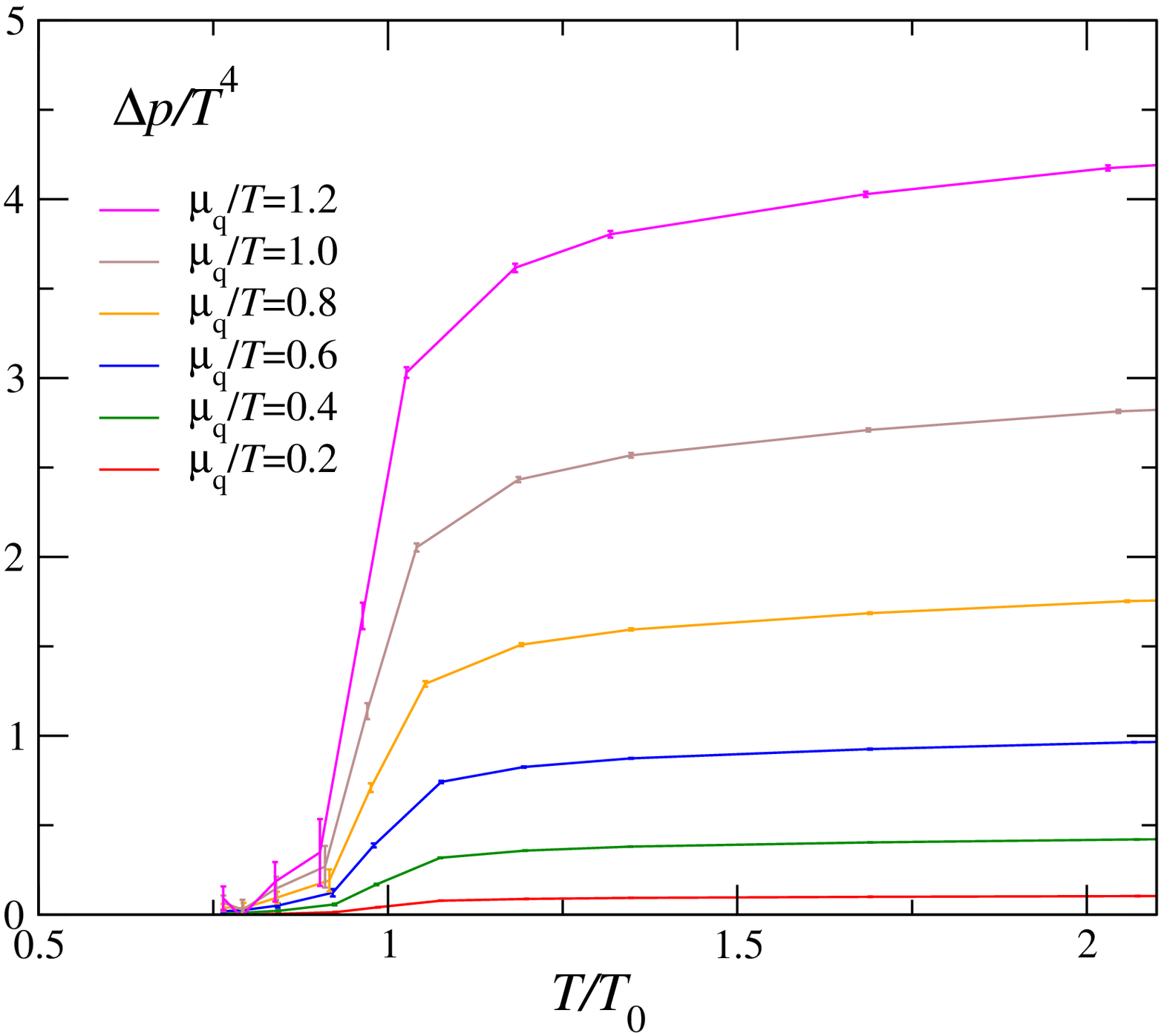}
\hspace{1mm}
\includegraphics[width=.32\textwidth]{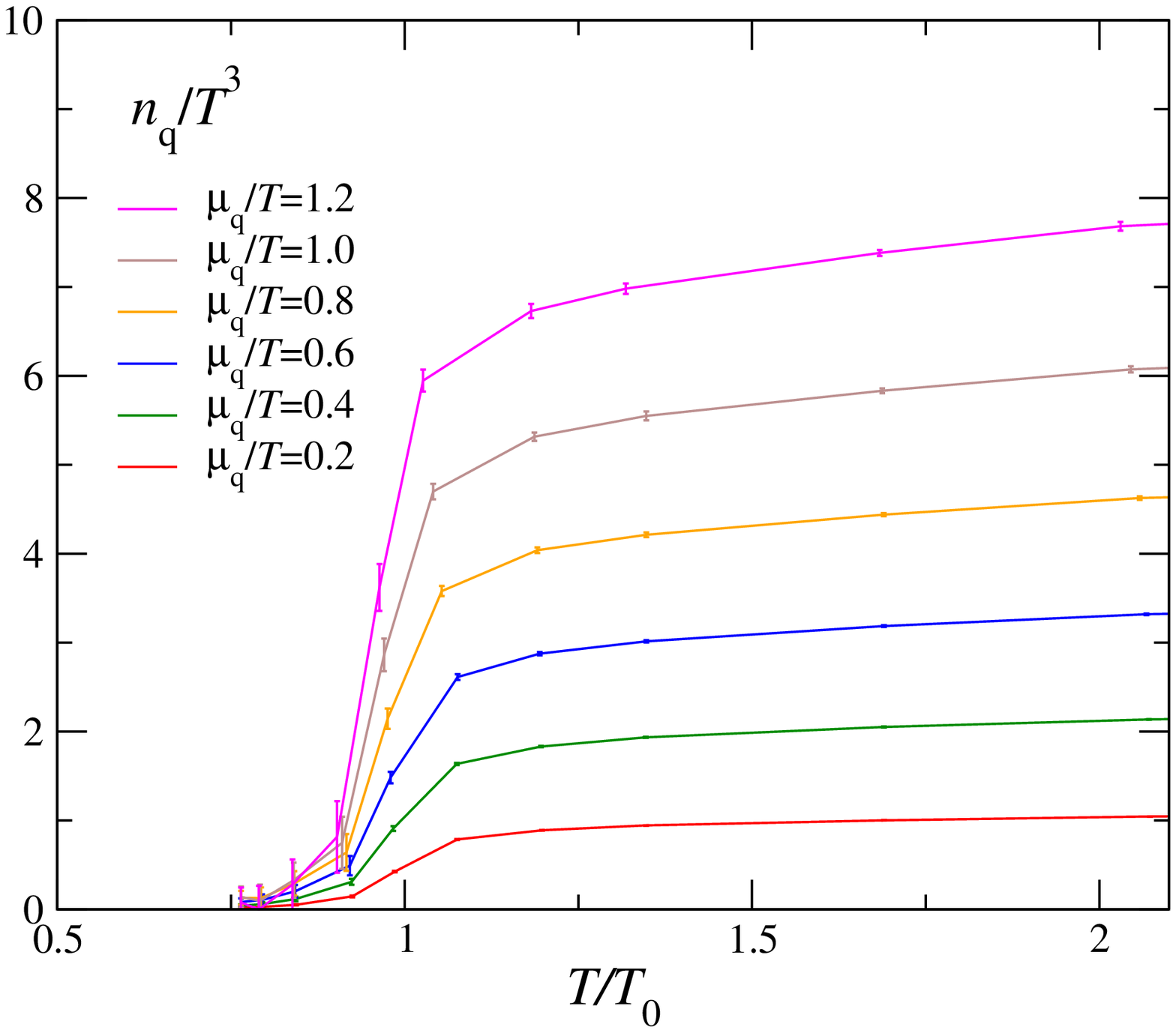}
\hspace{1mm}
\includegraphics[width=.32\textwidth]{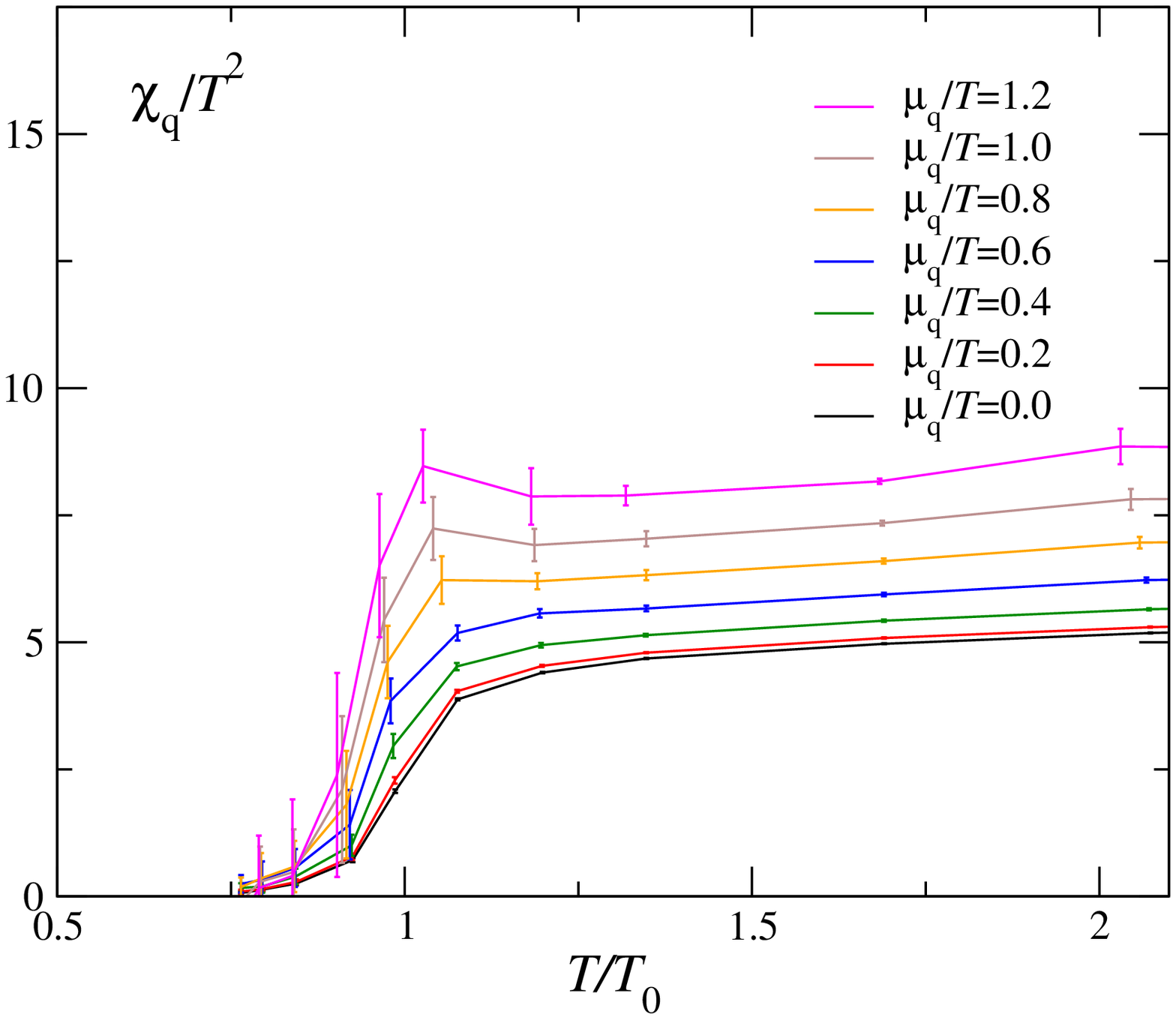}
}
\caption{Results of EOS using the hybrid method for $m_{\rm PS}/m_{\rm V} = 0.80$.}
\label{figHybrid80}
\end{figure}

Our results of EOS with these improvements are summarized in Figs.\ref{figHybrid65} and \ref{figHybrid80} for $m_{\rm PS}/m_{\rm V} = 0.65$ and 0.80, respectively.
Here, we calculate the quark number density and its susceptibility by numerical differentiations of the grand canonical
potential using the following thermodynamic formulae:
\begin{eqnarray}
\frac{n_q}{T^3} = \frac{N_t^3}{N_s^3} \,\frac{\partial (\ln {\cal Z})}{\partial (\mu_q/T)},
\hspace{5mm}
\frac{\chi_q}{T^2} = \frac{N_t^3}{N_s^3} \,\frac{\partial^2 (\ln {\cal Z})}{\partial (\mu_q/T)^2}.
\end{eqnarray}
We find that, in spite of the fact that the simulations at different $T/T_0$ are independent, the resulting EOS shown in these figures has smooth $T$- and $\mu_q$-dependence which is in accordance with theoretical expectations. 
Therefore, we think that the assumption ${\cal D}_n=0$ at $n > 4$ is well satisfied down to $T$ close to the transition temperature.
Furthermore, we find that the statistical fluctuations are much reduced over the results of the previous section (Fig.\ref{figTaylor}). 
This is due to the improvements including the Gaussian method for $\theta$-averaging and the $\beta$-reweighting for $F$-averaging.
The hybrid method with improvements provides us with a tractable way to calculate EOS with Wilson-type quarks.

Similar to the case of the previous section, the quark number susceptibility for $m_{\rm PS}/m_{\rm V} = 0.65$ show enhancement near $T_0$ with increasing $\mu_q$, although the statistical errors are not quite small yet. 
On the other hand, the quark number susceptibility for $m_{\rm PS}/m_{\rm V} = 0.80$ shown in Fig.\ref{figHybrid80} do not show rapid enhancement with $\mu_q$.
This may be in part explained by the expectation that the critical point locates at larger $\mu_q$ because the quark mass is larger than that for $m_{\rm PS}/m_{\rm V} = 0.65$.
The mild enhancement shown in Fig.\ref{figHybrid80} may be suggesting that, at this quark mass, the critical point does not locate inside the applicability range in $\mu_q$ with our ${\cal O}(\mu_q^4)$ calculation.
Further studies with increased statistics around $T_0$ are needed for more definite conclusions.

\section{Conclusions}

We have carried out the first calculation of the equation of state at non-zero densities with two flavors of improved Wilson quarks.  Statistical fluctuations of physical observables at finite density are much severer with Wilson-type quarks than with staggered-type quarks.  To tame the problem, we combined and developed several improvement techniques.

With these improvements, we found that the peak height of the quark number fluctuation at the pseudo-critical temperature increases as $\mu_q$ increases.  
In contrast, isospin susceptibilities show no sharp peaks at the pseudo-critical temperature.  
These results agree with previous observations by the Bielefeld-Swansea Collaboration using staggered-type quarks, and suggest that a critical point exists at finite $\mu_q$, which is expected to locate at the end point of a first order transition line between confining and deconfining phases in the coupling parameter space of $T$ and $\mu_q$.
Details of our study will be described in \cite{paperWHOT}.

\vspace{5mm}   
This work is in part supported by Grants-in-Aid of the Japanese Ministry
of Education, Culture, Sports, Science and Technology
(Nos. 17340066, 18540253, 19549001 and 20340047). 
SE is supported by U.S.\ Department of Energy (DE-AC02-98CH10886).
Numerical calculations were performed on supercomputers at KEK by the Large Scale Simulation Program Nos. 06-19, 07-18, 08-10, at CCS, Univ.\ of Tsukuba, and at ACCC, Univ.\ of Tsukuba.


\end{document}